\documentclass[]{jfm}

\usepackage{graphicx}
\usepackage{newtxtext}
\usepackage{newtxmath}
\usepackage{natbib}
\usepackage{xcolor}
\usepackage{hyperref}
\hypersetup{
    colorlinks = true,
    urlcolor   = blue,
    citecolor  = black,
}

\newcommand{\RomanNumeralCaps}[1]
\linenumbers

\newcommand{\MB}[1]{{\color{black} #1}}

\newcommand{\JP}[1]{{\color{black} #1}}
\newcommand{\RK}[1]{{\color{black} #1}}

\title{Multistability of elasto-inertial two-dimensional channel flow}

\author{Miguel Beneitez\aff{1}
  \corresp{\email{mb2467@cam.ac.uk}},
  Jacob Page\aff{2}, Yves Dubief\aff{3}
 \and Rich R. Kerswell\aff{1}}

\affiliation{ \aff{1}DAMTP, Centre for Mathematical Sciences, Wilberforce Road, Cambridge CB3 0WA, UK 
\aff{2}School of Mathematics, University of Edinburgh, EH9 3FD, UK
\aff{3}Department of Mechanical Engineering, University of Vermont, Burlington, VT, USA}

\begin{document}
\maketitle

\begin{abstract}
Elasto-inertial turbulence (EIT) is a recently discovered two-dimensional chaotic flow state observed in dilute polymer solutions. It has been hypothesised that the dynamical origins of EIT are linked to a center-mode instability, whose nonlinear evolution leads to a travelling wave with an `arrowhead' structure in the polymer conformation, a structure also observed instantaneously in simulations of EIT. In this work we conduct a suite of two-dimensional direct numerical simulations spanning a wide range of polymeric flow parameters to examine the possible dynamical connection between the arrowhead and EIT. Our calculations reveal (up to) four co-existent attractors: the laminar state and a steady arrowhead, along with EIT and a `chaotic arrowhead'. The steady arrowhead is stable for all parameters considered here, while the final pair of (chaotic) flow states are visually very similar and can be distinguished only by the presence of a weak polymer arrowhead structure in the `chaotic arrowhead' regime. Analysis of energy transfers between the flow and the polymer indicates that both chaotic regimes are maintained by an identical near-wall mechanism and that the weak arrowhead does not play a role. Our results suggest that the arrowhead is a benign flow structure that is disconnected from the self-sustaining mechanics of EIT.
\end{abstract}


\section{Introduction} 
It has been more than 70 years since the phenomenon of polymer drag reduction in wall-bounded turbulence was first observed experimentally \citep{toms1948some,mysels1949}.
Following this discovery, great efforts have been directed towards understanding how inertial turbulence (IT) is altered by the addition of polymers to the flow \citep[e.g. see the reviews][]{lumley1969drag,white2008mechanics}. Polymeric fluids also exhibit counter-intuitive chaotic behaviour in very small scale, inertialess flows.
This `elastic' turbulence (ET) was also first discovered experimentally \citep{groisman2000elastic,groisman2004elastic} at vanishing Reynolds numbers and is thought to rely on finite-amplitude curvature in the streamlines \citep{shaqfeh1996purely}.
In contrast to polymer-modified IT, ET is associated with an increased drag relative to the laminar Newtonian state \citep{varshney2018drag}.
It can be exploited to promote heat transfer \citep{traore2015efficient} and to efficiently mix at very small scales \citep{squires2005microfluidics}.

In addition to these distinct phenomena, a third chaotic flow state was recently identified \citep{samanta2013elasto,dubief2013mechanism} where both inertial and elastic effects are relevant, and was named `elasto-inertial' turbulence (EIT). 
EIT can be sustained for Reynolds numbers $Re = O(1000)$, and is potentially linked to the `early turbulence' reported in a range of experimental studies \citep{jones1966onset,goldstein1969turbulent,draad1998laminar,choueiri2018exceeding,chandra2018onset}.
EIT differs from both IT and ET in that it can be sustained in a purely two-dimensional planar flow \citep{sid2018two}, and is dominated by highly extended `sheets' of polymer stress \citep[e.g. see][]{dubief2023elasto}.
A connection has been sought between EIT and the so-called `maximum drag reduction' state in IT \citep{zhu2021nonasymptotic,zhang2021role}, though the mechanisms underpinning both of these flow types remain to be clarified.

Despite much progress in our statistical understanding of the various chaotic viscoleastic flows \citep{datta2022perspectives,sanchez2022understanding,dubief2023elasto}, the dynamical origins and connections between polymer-perturbed IT, EIT and ET remain largely unknown.
The exception here is ET in curved geometries, which is associated with a linear instability driven by viscoelastic hoop stresses \citep{larson1990purely,shaqfeh1996purely}.
In parallel flows there has been some indication that self-sustaining ET can be triggered by a finite amplitude perturbation to generate the curvature necessary for a hoop-stress instability \citep{meulenbroek2004weakly,morozov2007introductory,pan2013nonlinear}, but the exact requirements and dynamical connection to the linear instabilities in a curved geometry has not been demonstrated and
there is also the possibility of a direct connection to EIT.

The situation in a planar pressure-driven channel flow is ripe for investigation due to the presence of a pair of linear instabilities.
One is the viscoelastic analogue of the Newtonian Tollmien-Schlichting (TS) waves and exists at high $Re$ \citep{zhang2013linear}. 
It has been observed that the polymer conformation field associated with a saturated TS wave and the polymer conformation for the weakly chaotic edge state for (subcritical) EIT have a similar appearance \citep{shekar2019critical,shekar2021tollmien} though the TS branch turns around prior to the emergence of EIT as the Weissenberg number $Wi$ is increased and a clear dynamical connection has yet to be established.
The other instability was discovered only very recently, and is a `centre mode' found in both pipes \citep{garg2018viscoelastic} and channels \citep{Khalid2021a} at modest Weissenberg numbers $Wi \sim 20$.
Most intriguingly, the unstable centre mode in a channel remains unstable even in the inertialess limit \citep{khalid2021b}, although only for very high $Wi$ and vanishing polymer concentration \citep[more realistic values of $Wi$ are found with the introduction of a more realistic polymer model, see][]{buza2022a}.
The existence of a linear instability in areas of the parameter space relevant to \emph{both} ET and EIT could provide a plausible direct connection between these states. 
The nonlinear evolution of the viscoelastic centre mode leads to a saturated `arrowhead' travelling wave \citep{page2020exact} which is strongly subcritical \citep{wan2021subcritical,buza2022a}.
The arrowhead can be continued down to the inertialess limit where it is found to exist at experimentally realisable values of the Weissenberg number \citep{buza2022b,morozov2022coherent}.
Finite amplitude structures which are similar in appearance to the exact arrowhead travelling waves have been observed in experiments at low $Re$ \citep{choueiri2021experimental} and have also been seen intermittently in numerical simulations of EIT at high $Re$ \citep{page2020exact,dubief2020first}.
However -- much like the TS waves -- a direct route to chaos from this structure (e.g. in a sequence of successive bifurcations) has yet to be found. 

The possible importance of the arrowhead in sustaining EIT was suggested by the simulations of \citet{dubief2020first}, who
performed DNS of viscoelastic flows using the FENE-P model for $Re=1000$, $Wi \in [50,200]$ and $0.5\leq \beta \leq 1$. 
Their study identified several regimes in different areas of the parameter space: 
a stable travelling wave arrowhead, EIT, a chaotic arrowhead and an intermittent arrowhead. 
Motivated by these results, we conduct a systematic study of the state space of a two-dimensional viscoelastic channel flow for a wide range of polymeric parameters in the FENE-P model, in an effort to directly connect the arrowhead to EIT. 
Surprisingly, we find that the arrowhead is a benign flow structure -- it can be maintained on top of a background EIT, but does not play a role in the self-sustaining mechanism which is driven by near-wall behaviour. 
Our search reveals that the steady arrowhead travelling wave is always stable for the parameters we consider, and we also find a large region of multistability with up to four attractors -- the laminar state, a steady arrowhead, EIT and a chaotic arrowhead.
The final regime is nearly identical to EIT apart from a weak arrowhead in the centre of the domain. 

The rest of this paper is structured as follows:
In \S 2 we present the governing equations and describe the numerical simulations to be conducted.
In \S 3 we present evidence for the four distinct attractors and draw connections to the results of \citet{dubief2020first}.
In \S 4 we look for dynamical connections between the attractors and compute various edge states between them. Finally, conclusions are presented in \S 5. 

%
%
\section{Formulation and computational details}
\label{sec:formulation}
We consider two-dimensional streamwise-periodic flow between two infinite, stationary, rigid walls, separated by a distance $2h$ and driven by a time-varying pressure-gradient so that the mass flux is constant. The viscoelastic flow is governed by the finite-extensibility nonlinear elastic-Peterlin (FENE-P) model with governing equations 
\begin{align}
    \partial_t \mathbf{u} + (\mathbf{u}\cdot \nabla)\mathbf{u} + \nabla p &= \frac{\beta}{\textit{Re}}\Delta \mathbf{u}+\frac{(1-\beta)}{\textit{Re}}\nabla \cdot \mathbf{T}(\mathbf{C}),\label{eq:Ueq}\\
    \partial_t \mathbf{C}+ (\mathbf{u}\cdot \nabla ) \mathbf{C} + \mathbf{T}(\mathbf{C}) &= \mathbf{C}\cdot \nabla \mathbf{u} + (\nabla \mathbf{u})^T\cdot \mathbf{C}+ \frac{1}{\textit{Re}\ \textit{Sc}}\Delta \mathbf{C}, \label{eq:Ceq} \\
    \nabla \cdot \mathbf{u} &= 0, \label{eq:divFree}
\end{align}
where
\begin{align}
    \mathbf{T}(\mathbf{C}) := \frac{1}{\textit{Wi}}\left(f(\text{tr} \mathbf{C})\mathbf{C}-\mathbf{I}\right),\quad \text{and}\quad f(x):=\left(1-\frac{x-3}{L_{\text{max}}^2}\right)^{-1}.
\end{align}
We consider two-dimensional flows with $\mathbf{u}=(u,v)$ denoting the streamwise and wall-normal velocity components, $p$ the pressure and $\mathbf{C}$ the positive-definite conformation tensor which represents the ensemble average of the product of the end-to-end vector of the polymer molecules. 
The parameter $\beta:=\nu_s/(\nu_s + \nu_p)$ denotes the viscosity ratio, with $\nu_s$ and $\nu_p$ the solvent and polymer contributions to the total kinematic viscosity, $\nu= \nu_s + \nu_p$
The parameter $L_{\text{max}}$ is the maximum extensibility of the polymer chains. The equations are made non-dimensional with the half-distance between the plates $h$ and the bulk velocity
\begin{equation}
    U_b:=\frac{1}{2h}\int_{-h}^{h}u\ dy.
\end{equation}
The non-dimensional Reynolds, \textit{Re}, and Weissenberg, \textit{Wi} numbers are defined as
\begin{equation}
    \textit{Re}:=\frac{U_bh}{\nu} \quad \text{and} \quad \textit{Wi}:=\frac{\tau U_b}{h},
\end{equation}
where $\tau$ denotes the polymer relaxation time. 
Equation ~\eqref{eq:Ceq}
has a stress diffusion term which for a realistic polymer solution would take a value $\textit{Sc} = O(10^{6})$. 
However, numerical simulations are typically restricted to much smaller values, $Sc=O(10^3)$ \citep{sid2018two,page2020exact}, and the term itself is treated as regulariser to help maintain positive-definite $\mathbf C$.
With non-zero polymer diffusion we must specify boundary conditions on the polymer conformation.
We apply the following equation at the wall:
\begin{equation}
    \partial_t \mathbf{C}+ \mathbf{T}(\mathbf{C}) = \mathbf{C}\cdot \nabla \mathbf{u} + (\nabla \mathbf{u})^T\cdot \mathbf{C}+ \frac{1}{\textit{Re}\ \textit{Sc}}\partial_{xx} \mathbf{C}, \label{eq:bcC}
\end{equation}
as previously used in the literature \citep{dubief2020first,page2020exact,buza2022b}. Note that equation ~\eqref{eq:Ceq} does not require boundary conditions in the limit $\textit{Sc}\to\infty$ \citep{sid2018two}; the boundary conditions \eqref{eq:bcC} are chosen so that the distance from the $\textit{Sc}\to\infty$ limit is minimized \citep{sid2018two,dubief2020first}. 

\subsection{Numerics}

The spectral codebase Dedalus \citep{burns2020dedalus} is used to perform direct numerical simulations of equations ~\eqref{eq:Ueq}-\eqref{eq:divFree}. 
We consider a computational domain of fixed size $[L_x,L_y]=[2\pi,2]$ in units of $h$. 
The quantities $\mathbf{C}$ and $\mathbf{u}$ are expanded in $N_x$ Fourier modes in the $x$ direction and $N_y$ Chebyshev modes in the $y$ direction. 
Time integration is performed with a 3rd-order semi-implicit BDF scheme \citep{wang2008variable} with fixed time step. 
We fix $\textit{Sc}=500$ for the majority of simulations unless otherwise indicated. 
The different numerical solutions have various requirements in term of resolution. We typically use $[N_x,N_y]=[512,600]$ to simulate travelling waves while higher values of $[N_x,N_y]=[600,800]$ are used to simulate chaotic states, although for some of the \MB{higher Weissenberg and higher $L_{\text{max}}$} cases we have considered $[N_x,N_y]=[800,1024]$. 
Further increasing $\textit{Sc}$ can cause $\mathbf{C}$ to lose positive definiteness in several locations of the domain, as previously reported \cite{dubief2020first}. 
However, reducing the computational timestep \MB{and increasing the resolution can alleviate this. We have checked those results which temporarily lose positive definiteness in certain regions by reducing the time step below $10^{-5}$ and increasing the resolution to at least $[N_x,N_y]=[2048,2048]$. The increase of spatio-temporal resolution ensured that positive definiteness is recovered while the reported dynamics remain unaltered.}

\subsection{Elasto-inertial attractors} 
\citet{dubief2020first} identified various statistically-steady states in the same channel geometry:
the laminar state (L), steady arrowhead (SAR), EIT and a chaotic arrowhead (CAR).
Note that \cite{dubief2020first} also discuss an intermittent arrowhead state (IAR) which we now believe is actually a weaker version of CAR and not a distinct state.

%
%
\begin{figure}
    \centering
    \includegraphics[width=0.8\textwidth]{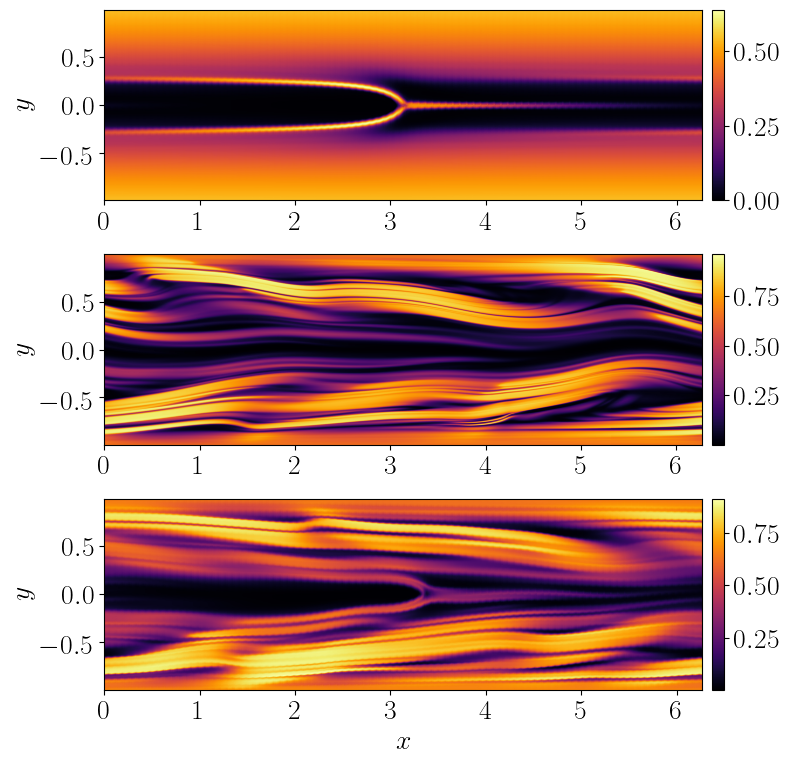}\put(-300,295){$(a)$}\put(-300,200){$(b)$}\put(-300,105){$(c)$}
    \caption{Example snapshots of tr$({\bf{C}})/L^2_{\text{max}}$ for the states initially explored in \cite{dubief2020first}. (a) Steady arrowhead regime (SAR) \MB{at $\textit{Re}=1000$, $\textit{Wi}=50$, $\beta=0.9$, $L_{\text{max}}=90$, $\textit{Sc}=500$}, (b) Elasto-inertial turbulence (EIT) \MB{at $\textit{Re}=1000$, $\textit{Wi}=50$, $\beta=0.9$, $L_{\text{max}}=70$, $\textit{Sc}=500$}, (c) Chaotic arrowhead regime (CAR) \MB{at $\textit{Re}=1000$, $\textit{Wi}=50$, $\beta=0.9$, $L_{\text{max}}=70$, $\textit{Sc}=500$}. We will show that these state do not succeed each other but coexist in parameter space. }
    \label{fig:3states}
\end{figure}
Examples of the the three non-trivial attractors are reported in figure \ref{fig:3states}, where we show contours of the polymer trace $\text{tr}({\mathbf C}) / L_{\text{max}}^2$.
The SAR state features a pair of symmetric sheets of polymer extension, which sit close to the channel centreline, bending to meet at $y=0$. 
A highly stretched central sheet then extends downstream along the centreline for almost half of the computational domain.
Both EIT and CAR show intense polymer stretch in near wall regions, with many wavy sheets of polymer extension layered on top of one another. 
The states are visually very similar, though CAR features a weak, distorted arrowhead structure near the centre of the domain.

We trigger each of the states discussed above and shown in figure \ref{fig:3states} by time-stepping appropriate initial conditions.
The SAR attractor was initially found via nonlinear saturation of the linear centre mode instability as described in \cite{page2020exact}. 
We found the SAR to always be stable, and were able to obtain this state at other parameter settings by supplying a converged arrowhead obtained nearby in parameter space as an initial condition.
We triggered the chaotic states CAR and EIT by applying blowing and suction at the wall, starting from either SAR (to obtain CAR) or the laminar state (to obtain EIT). 
The blowing and suction is \MB{similar} to that used in previous studies \citep{samanta2013elasto,dubief2013mechanism} and takes the form
\begin{equation}
    v(y=\pm 1) = \mp A\sin{(2\pi x/L_x)},
\end{equation}
with $A=2\times10^{-3}$. 
The forcing is active for $0 \leq t < 3$. 
Perturbations from the wall were found to be necessary to trigger the self-sustained chaotic states. 
In contrast, arbitrary perturbations applied in the core of the domain did not trigger chaotic behaviour. 

%
%
\section{Multistability of two-dimensional viscoelastic channels}

In this section we summarise our computations and map out regions of multistability. We also explore the impact of changing the flow parameters on the appearance and statistical properties of the various attractors.

%
%
\begin{figure}
    \centering
    \includegraphics[width=0.9\textwidth]{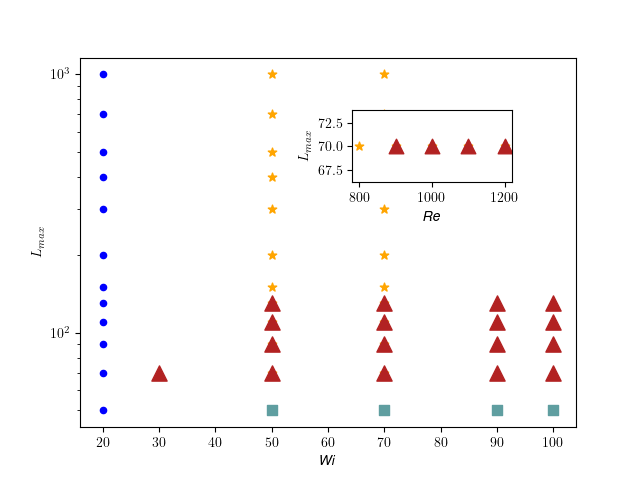}
    \caption{\JP{Summary of computations and the attractors found over parameter space.} 
    \RK{Blue circles indicates only the laminar (L) state was found as an attractor, orange stars  - the steady arrowhead (SAR) and L coexist as attractors, light blue squares - L and EIT co-exist as attractors, and red triangles - L, EIT, SAR and the chaotic arrowhead (CAR) all co-exist as attractors.  For the main plot, $Sc=500$, $\beta=0.9$ and $Re=1000$ while for the inset $Wi=50$, $\beta=0.9$ and  $Sc=500$ again. } \MB{At $L_{\text{max}}=50$ only EIT and L were explored as attractors as SAR/CAR become prohibitively expensive computationally.}}
    \label{fig:parReg}
\end{figure}

%
%
\subsection{Coexistence of attractors in parameter space}

The \RK{parameter space here is 5-dimensional and so a systematic search was impractical. However, a preliminary investigation indicated that $Wi$ and $L_{\text{max}}$ were the most important parameters (yielding the most qualitative changes) so they were the focus of the search: see figure \ref{fig:parReg}. Over the range $(Wi,L_{\text{max}}) \in [20,70] \times [50,1000]$ at $Sc=500$, $\beta=0.9$ and $Re=1000$, the laminar state is linearly stable. This is consistent with the centre-mode mode instability appearing at slightly higher $Wi$ (see figure 2 in \cite{page2020exact} where $Sc=1000$  was used). However, the consequence of the centre-mode instability - the SAR state - is seen at lower $Wi$ as the instability is subcritical.  The SAR state was found to be an attractor for $L_{\text{max}} \geq 70$ and $Wi \geq 30$ consistent with \cite{page2020exact}. 
Interestingly,  the chaotic arrowhead state (CAR) was only found where SAR also exists and is stable (basically for $Wi \geq 50$ and $L_{\text{max}} \in [70,120]$) excluding the possibility of a SAR-to-CAR bifurcation in this $(Wi, L_{\text{max}})$ range. EIT was found for $\textit{Wi}\in[30,100]$, $L_{\text{max}}\in[50,130]$ with  $\beta\in[0.9,0.97]$, $\textit{Re}\in[900,1200]$ and $Sc \geq 500$. In terms of fig \ref{fig:parReg}, CAR and EIT coexist when  $L_{\text{max}}$ is as low as $50$ where only EIT and the laminar state 
\JP{were simulated.} 
\JP{Attempts to simulate} \MB{SAR and CAR for this value of $L_{\text{max}}$ were prohibitively expensive computationally due to the loss of positive definiteness and high spatio-temporal resolution required. } }

\RK{The general conclusion from figure \ref{fig:parReg} is that the nonlinear states firstly reported in \citep{dubief2020first} - SAR, CAR \& EIT - coexist in parameter space rather than succeeding each other as attractors. The latter scenario would suggest dynamical connections between the states in which one loses stability to another but this seems not to be the case at least in the parameter ranges considered. The fact that SAR and CAR coexist as attractors for the parameters considered is particularly surprising as CAR  plausibly looks like the indirect result of a bifurcation off SAR.}

%
%
\begin{figure}
    \centering
    \includegraphics[width=0.47\textwidth]{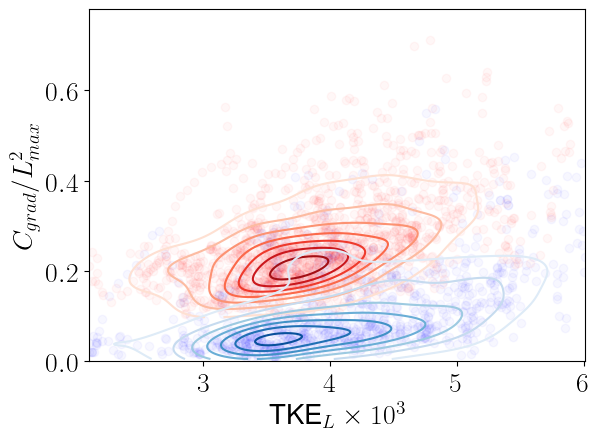}\put(-180,130){$(a)$}
    \includegraphics[width=0.47\textwidth]{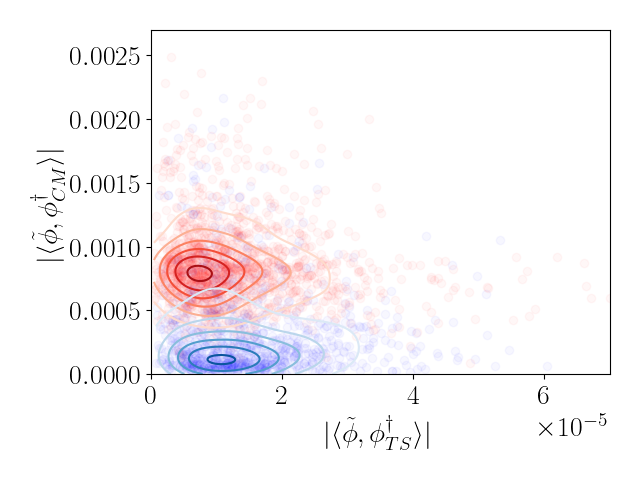}\put(-170,130){$(b)$}\\
    \includegraphics[width=0.47\textwidth]{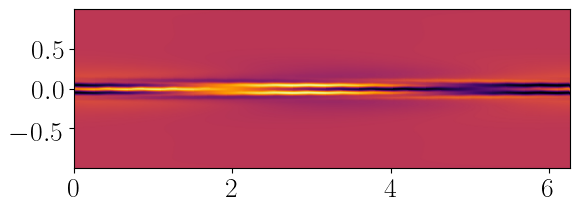}\put(-180,70){$(c)$}
    \includegraphics[width=0.47\textwidth]{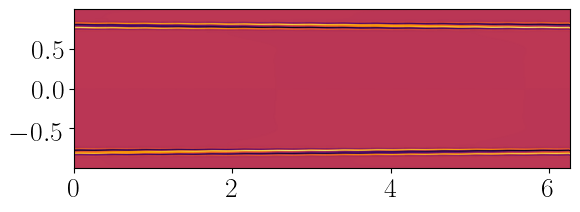}\put(-180,70){$(d)$}
    \caption{\MB{(a) $C_{\text{grad}}/L^2_{\text{max}}$ \textit{vs.} $TKE_L$ as defined in the main text for for EIT (red) and CAR (orange) identified at \textit{Re}=1000, \textit{Wi}=50, $L_{\text{max}}=70$, $\beta=0.9$, $\textit{Sc}=500$ for a finite time interval $T\approx 1000$. (b) Projection of the same EIT trajectory (red) and CAR (orange) onto the TS mode \textit{vs.} the projection onto the centermode. The figures present observables to show that EIT and CAR are two separate attractors. (c) trace of the centermode for the aforementioned parameters and $k_x=1$. (d) idem for the TS mode which becomes unstable at sufficiently large \textit{Re}. Note that the projection of the TS mode is much smaller than that of the centermode due to the smaller spatial extension of $\text{tr}(\bf{C})$, which is the largest term in the corresponding eigenmode.}}
    \label{fig:statePort}
\end{figure}

%
%
\RK{\subsection{Distinquishing between CAR and EIT \label{3.2}}}

\RK{Figure \ref{fig:3states} shows that the CAR and EIT states look very similar and developing some quantitative measure to distinquish them is important. Another issue is whether either state is just a long-lived transient. For example, does CAR eventually evolve into the EIT state? This latter question is impossible to answer definitively with finite-time computations but what can be said is that over the course of our simulations (some of duration over 1000 $h/U_b$), CAR never collapsed. }

\RK{The defining feature of CAR is the mixture of an arrowhead  structure at the midplane with the chaotic stretched polymer sheets towards the walls which characterise EIT. A quantity well suited to picking the former feature out is 
\begin{equation}
    C_{\text{grad}} := \frac{1}{L_x}\int |\partial_x C'_{kk}(x,y=0)| dx,
\end{equation}
which is the streamwise-averaged gradient magnitude  of the perturbation \MB{over the laminar state} trace along the centreline. We also use the $L_2$-norm of the velocity difference from the laminar flow - a turbulent kinetic energy
\begin{equation}
    \text{TKE}_{L}:=\frac{1}{2L_{x}}\int_{\Omega} {\bf (u-u_L)}^2 d\Omega, \label{eq:TKE_L}
\end{equation}
to compare CAR and EIT.  Figure \ref{fig:statePort}~(a) shows the two-dimensional probability density function (PDF) over these two quantities for the CAR and EIT states collected over a 1000 $h/U_b$ time period. The turbulent kinetic energy of EIT and CAR are very similar but, as expected, $C_{\text{grad}}$ is much larger for CAR than EIT.}

\RK{Another differentiator between EIT and CAR} is the result of  projecting onto the eigenmodes of the symmetric centre-mode (CM) and the antisymmetric Tollmien-Schlichting (TS) mode as follows
\begin{equation}
    \langle \,\phi^{\dagger}_{j}, \phi \, \rangle = \frac{1}{2} \int_{-1}^{1} \phi^{\dagger *}_{j}    \phi \,dy, \label{eq:emode_proj}
\end{equation}
with
\begin{equation}
    \phi(y):=\frac{1}{L_x}\int_{0}^{L_x}\varphi(x,y) e^{i x}dx,
\end{equation}
\MB{where $j=\{\text{CM},\text{TS}\}$}, $\varphi=[u',v',p',C'_{xx},C'_{yy},C'_{zz}, C'_{xy}]$ is the perturbation \RK{to the laminar state }, and $\phi$ denotes the projection onto the $k_x=1$ mode ($^*$ denotes complex conjugate and $^\dagger$ the adjoint). \RK{Figure~\ref{fig:statePort}~(b) shows that this projection for the same perturbation trajectories used in figure~\ref{fig:statePort}~(a) produces the same desired separation.}
The projection onto the centre-mode is much larger than the TS mode one for both chaotic states. \MB{This is caused by the fact that the trace of the conformation tensor $\bf C$ in the TS eigenmodes has a much larger amplitude than the other components, but its spatial extension is significantly smaller: see figure~\ref{fig:statePort}~(c) and (d).}

%
%
\subsection{Effect of varying $Wi$ and $L_{\text{max}}$} 

%
%
\begin{figure}
    \centering
    \includegraphics[width=0.7\textwidth]{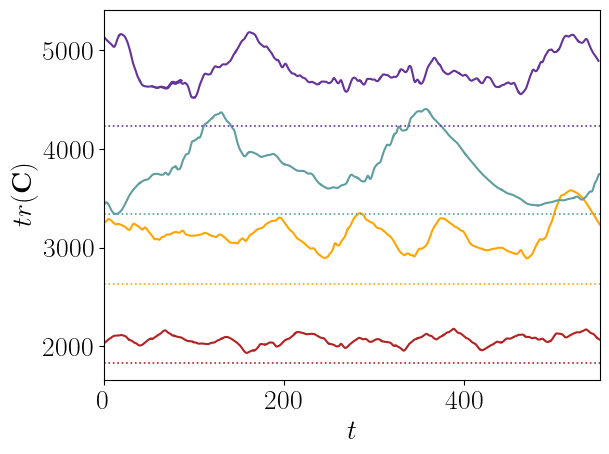}
    \caption{(a) Timeseries corresponding to several chaotic arrowheads (solid) and the corresponding SAR (dashed) at \textit{Re}$=1000$, \textit{Wi}$=50$, $\beta=0.9$ with $Sc=500$ for $L_{\text{max}}=\{110,90,70\}$ (second top to bottom) and $Sc=1000$ for $L_{\text{max}}=130$ (top). \MB{The figure shows how the duration of the calm-active phases becomes longer with increasing $L_{\text{max}}$, i.e. the peaks of tr$(\bf{C})$ become \RK{more separated} in time. This shows that the intermittent arrowhead regime (IAR) and CAR reported in \cite{dubief2020first} are }\RK{smoothly connected and so correspond to the same attractor.}}
    \label{fig:CAH_timeseries}
\end{figure}

The kinetic energy of the steady arrowhead state, SAR, increases for increasing $L_{\text{max}}$, in line with the results previously reported \citep{dubief2020first,buza2022b}.
Figure~\ref{fig:CAH_timeseries} (a) shows the time series of the volume-averaged trace for CAR corresponding to $L_{\text{max}}=\{70,90,110\}$ \RK{at $(Wi,Re,Sc,\beta)=(50,1000,500,0.9)$ and $L_{\text{max}}=130$ at $(Wi,Re,Sc,\beta)=(50,1000,1000,0.9)$} where the change in $Sc$ was necessary to maintain chaotic dynamics (ditto for the corresponding SAR). The figure shows that the CAR states undergo periods of calmer, less energetic dynamics alternating with more active periods. The duration of the calmer events increases with $L_{\text{max}}$ \MB{as shown by the increasing distance between peaks of the time series}. This behaviour indicates a continuous transition between the previously reported chaotic arrowhead  and intermittent arrowhead regimes, leading to the conclusion that these states are two ends of the same attractor, hereafter labelled CAR. The intermittent arrowhead state \RK{discussed in \cite{dubief2020first} is simply a CAR state where the calm phases dominate the chaotic dynamics which occurs as $L_{\text{max}}$ gets large for example.}

\begin{figure}
    \centering
    \begin{tabular}{cc}
       \includegraphics[width=0.47\textwidth]{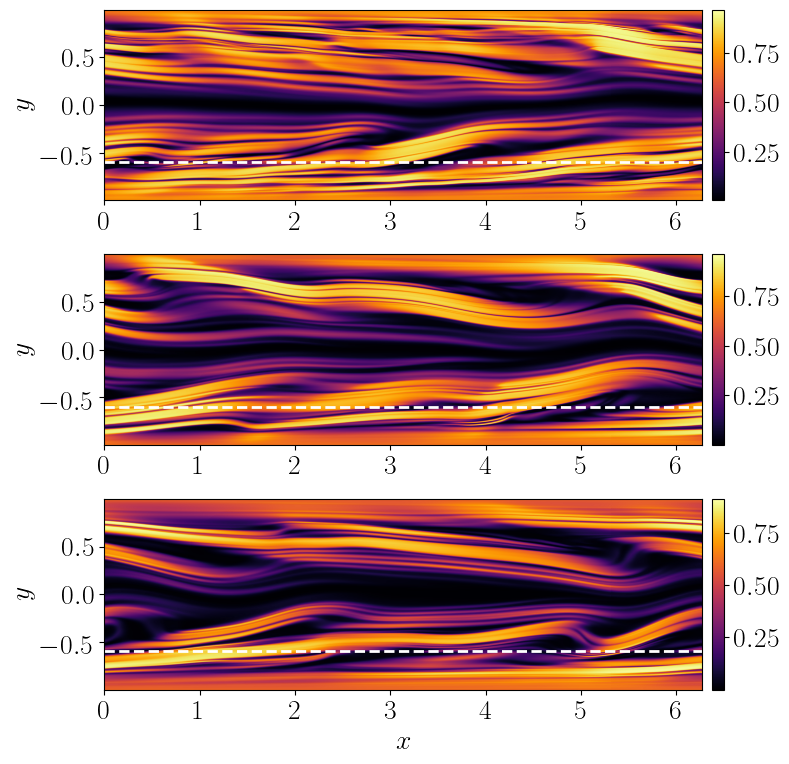} \put(-180,175){$(a)$}\put(-180,120){$(b)$}\put(-180,60){$c)$} & \includegraphics[width=0.45\textwidth]{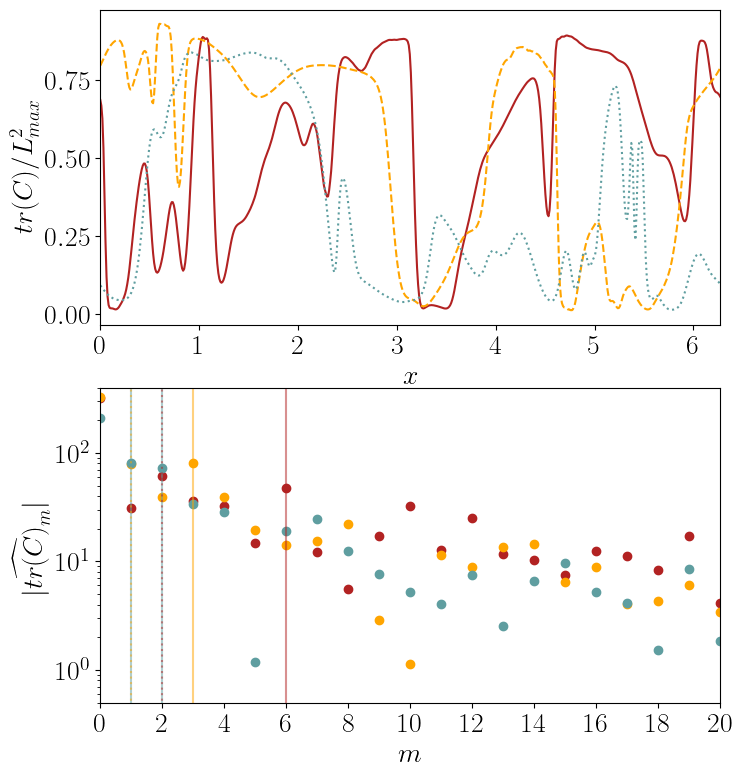} \put(-180,185){$(d)$}\put(-180,90){$(e)$}
    \end{tabular}
    \caption{Left: Snapshots of tr$({\bf{C}})/L^2_{\text{max}}$ of EIT with varying $L_{\text{max}}$ for fixed \textit{Re}$=1000$, \textit{Wi}$=50$, $\beta=0.9$, (a) $L_{\text{max}}=50$. (b) $L_{\text{max}}=70$. (c) $L_{\text{max}}=90$ and $Sc=500$ for all cases. Right: (d) tr$({\bf{C}})/L^2_{\text{max}}$ along the \MB{arbitrarily chosen} line $y=-0.6$ \MB{for $L_{\text{max}}=50$ (red), $L_{\text{max}}=70$ (orange), $L_{\text{max}}=90$ (blue)}. (e) Fourier transform of $L_{\text{max}}$ for the lines in the top right figure illustrating how the lengthscales in the flow increase with $L_{\text{max}}$, \MB{i.e. smaller $L_{\text{max}}$ shows greater amplitudes in the \RK{lower}  wavenumber modes. The vertical lines indicate the wavenumbers corresponding to the two \RK{smallest} wavenumbers (apart from 0) for each $L_{\text{max}}$ above.}}
    \label{fig:EITevol}
\end{figure}

\MB{The effect of \RK{varying} $L_{\text{max}}$ on the EIT states can be seen in figure~\ref{fig:EITevol} \RK{where  the  lengthscales in instantaneous snapshots } increase with $L_{\text{max}}$. This is further supported by considering $\text{tr}({\bf C})$ at any arbitrary horizontal line ($y=-0.6$ in this case), which is shown in Figure \ref{fig:EITevol}(d) and its Fourier transform in Fig \ref{fig:EITevol}(e). The latter figure shows that for \RK{in}creasing $L_{\text{max}}$, the energy content in the \RK{larger} scales (\RK{smaller} wavenumbers) is increased.}

%
%

%
%
\begin{figure}
    \centering
    \includegraphics[width=0.47\textwidth]{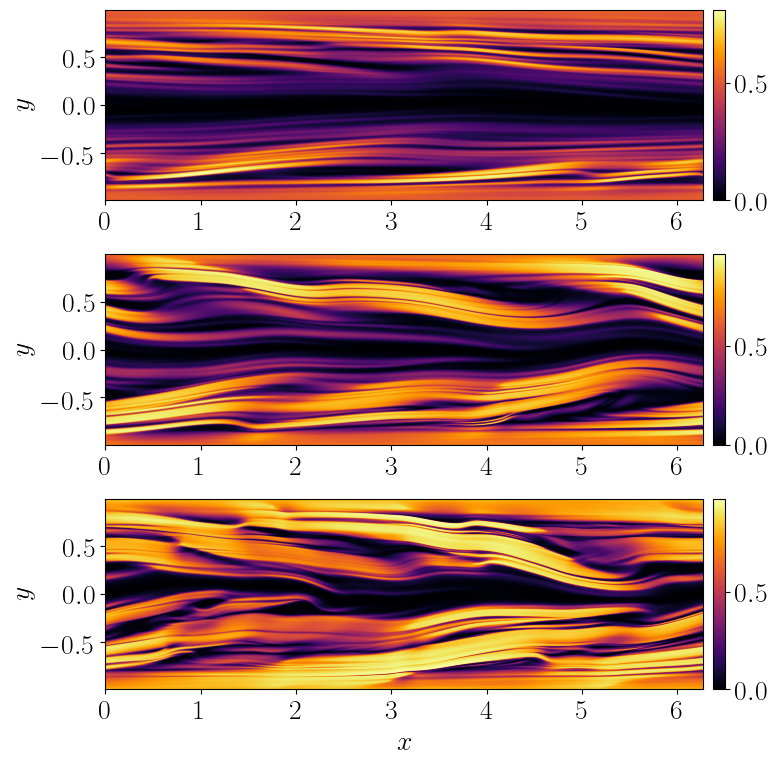}\put(-180,175){$(a)$}\put(-180,120){$(b)$},\put(-180,60){$(c)$}
    \caption{Snapshots of tr$({\bf{C}})/L^2_{\text{max}}$ of EIT with varying \textit{Wi} for fixed \textit{Re}$=1000$, $L_{\text{max}}=70$, $\beta=0.9$, \textit{Sc}$=500$. (a) \textit{Wi}=30. (b) \textit{Wi}=50. (c) \textit{Wi}=100.}
    \label{fig:EITevol_W}
\end{figure}

\RK{The effect of increasing $Wi$ on the EIT state is to make the polymer sheets more undulating spatially and temporally: see figure \ref{fig:EITevol_W}. Increasing $Wi$ also intensifies the polymer layers which reach closer to the centerline. }
The same trends were also observed for \RK{decreasing $L_{\text{max}}$ and are found also for the CAR state. A discussion about the evolution of the SAR states with \textit{Wi} can be found in \cite{buza2022b}}.

\RK{\subsection{Effect of varying $Re$, $\beta$ and $Sc$}}

\begin{figure}
    \centering
    \includegraphics[width=0.47\textwidth]{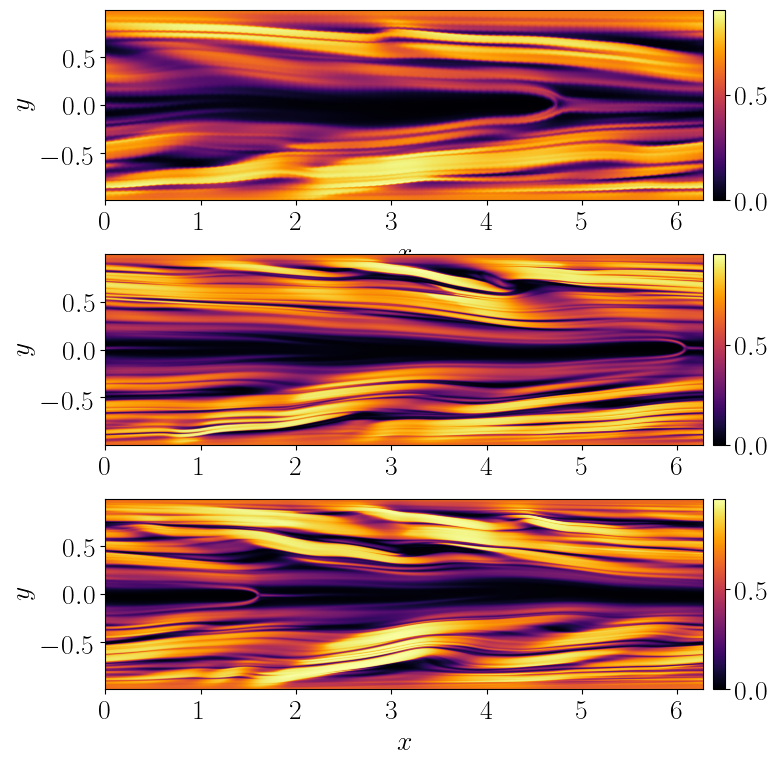}\put(-180,175){$(a)$}\put(-180,120){$(c)$}\put(-180,60){$(e)$}
    \includegraphics[width=0.47\textwidth]{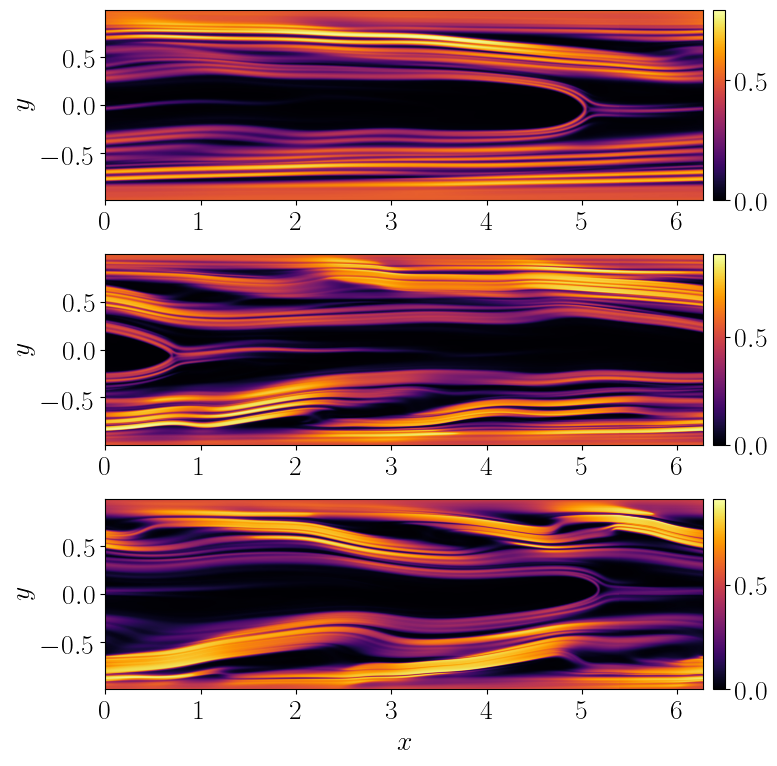}  \put(-180,175){$(b)$}\put(-180,120){$(d)$}\put(-180,60){$(f)$}

    \caption{Left: snapshots of tr$({\bf{C}})/L^2_{\text{max}}$ for CAR with varying $\beta$ at $\textit{Re}=1000$, $\textit{Wi}=50$, $L_{\text{max}}=70$ and \textit{Sc}$=500$: (a) $\beta=0.9$; (c) $\beta=0.95$; (e) $\beta=0.97$. 
    Right: snapshots of tr$({\bf{C}})/L^2_{\text{max}}$ for CAR with varying $\textit{Re}$ at $\textit{Wi}=50$, $L_{\text{max}}=120$, $\beta=0.9$ and \textit{Sc}$=500$: (b) $\textit{Re}=900$; (d) $\textit{Re}=1100$; (e) $\textit{Re}=1200$.
    \RK{Increasing  $\beta$ and $Re$ separately or together  intensifies the chaotic dynamics in agreement with \cite{dubief2020first}.}} 
    \label{fig:CAHevol_beta_Re}
\end{figure}
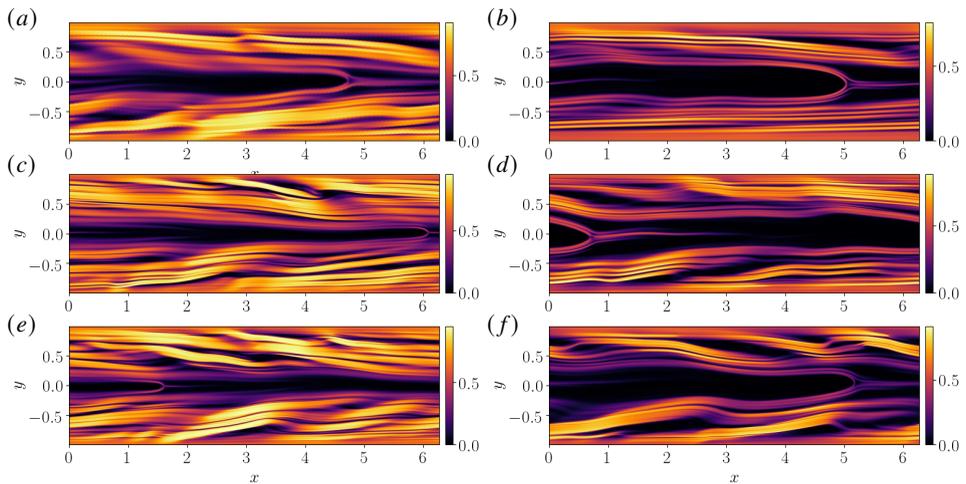

%
%
\RK{EIT and CAR remain robust as the  Reynolds number is increased away from where  they first  appear in parameter space}. As an example, figure~\ref{fig:CAHevol_beta_Re}(right) shows the CAR for three different $\textit{Re}=\{900,1100,1200\}$, while keeping the remaining parameters fixed. The intensity of the dynamics increases while the arrowhead persists at the centreline, consistent with \cite{dubief2020first}.

%
%

%
%
\RK{Increasing the polymer concentration, $\beta$, also intensifies the chaotic dynamics present}. Figure~\ref{fig:CAHevol_beta_Re}~(left) shows CAR for $\beta=\{0.9,0.95,0.97\}$ for fixed $\textit{Re}=1000,\ \textit{Wi}=50,\  L_{\text{max}}=70,\ \text{and}\ \textit{Sc}=500$. A larger $\beta$ leads to more active chaotic dynamics.
\RK{
Steady arrowhead states (SAR) have been observed at values as low as $\beta\approx 0.5$ \citep{dubief2020first,morozov2022coherent, buza2022b}, whereas we have found that  chaotic states (EIT and CAR) cannot be sustained for values below $\beta\approx 0.8$ (using $Re=1000$, $L_{\text{max}}=70$, $Wi=50$ and $Sc=500$)
}.

\RK{The majority of the results presented here were computed using $Sc=500$ as a compromise between including a vanishingly small real diffusion \citep[see e.g.][]{el1989existence} and enough diffusion to numerically stabilise the time-stepping spectral code at the resolutions used. The value of $Sc=500$ was also selected as the best match to the previous finite-difference computations reported in \cite{dubief2020first} and \cite{page2020exact} where a value of $Sc=1000$ was taken (finite difference codes already have some implicit numerical diffusion so less needs to be added explicitly to stabilise time-stepping compared to a spectral code). Even then, the EIT reported in fig 2 of \cite{page2020exact} (the red square at $Wi=20$, $\beta=0.9$, $L_{\text{max}}=500$ and $Re=1000$) could only be recovered by using neighbouring parameter values $Wi=30$, $\beta=0.9$, $L_{\text{max}}=120$ and $Re=1000$. Runs were also carried out with $Sc=150$ and 1000 which confirmed that all 4 states (EIT, CAR, SAR and L) as well as their coexistence are robust. The exact parameter limits for their coexistence, however, do depend on \textit{Sc} and taking $Sc \leq 50$ killed the chaotic states.}

%
%
\section{Dynamic connections between attractors}

\RK{The goal of this section is to explore how the various attractors -- L, SAR, CAR and EIT -- are organised in state space. Of primary concern is identifying which states share basin boundaries and which do not.}  The physical features present in the different states, such as the presence of a polymer sheet across the midplane or the undulations of polymer sheets closer to the wall, are common to several of the states identified. \RK{It is therefore natural to ask how transitions can occur between them and how they come into existence as the parameters are varied. }

\RK{As an initial check, we first examined the linear stability of the SAR state which results from  the \RK{centre-mode} instability found by \cite{garg2018viscoelastic} in a pipe and \cite{Khalid2021a} in a channel. This bifurcation is generally subcritical in both $Re$ and $Wi$ \citep{wan2021subcritical, buza2022a} with the steady arrowhead solution (SAR) emerging as the upper branch solution \citep{page2020exact, buza2022b, morozov2022coherent}}. 
We examined the two-dimensional linear stability of the SAR states performing a global stability analysis using an implicitly-restarted Arnoldi method \citep{sorensen1992implicit,bagheri2009matrix}. The linear stability analysis was carried out in the frame travelling with the speed of the SAR, where the state corresponds to a fixed point \RK{(the perturbation was represented by $N_x=64$ streamwise and $N_y=512$ wall-normal modes)}. All the SAR states tested were found to be linearly stable to 2-dimensional perturbations consistent with the time-stepping numerics. \RK{Interestingly, while this work was being performed, another group have found that the SAR state is, however, linearly unstable to 3-dimensional perturbations where there is a non-vanishing spanwise wavenumber \citep{lellep2023linear}.}

\RK{As the laminar state is also linearly stable over the parameter space being considered, the transition between the SAR, L and the other chaotic states must then be through finite amplitude perturbations. To shed some light on this, the saddle states lying in the boundaries between the basins of attraction of the aforementioned attractors, i.e. the edge states. are considered below.}

%
%
\begin{figure}
    \centering
    \includegraphics[width=1.0\textwidth]{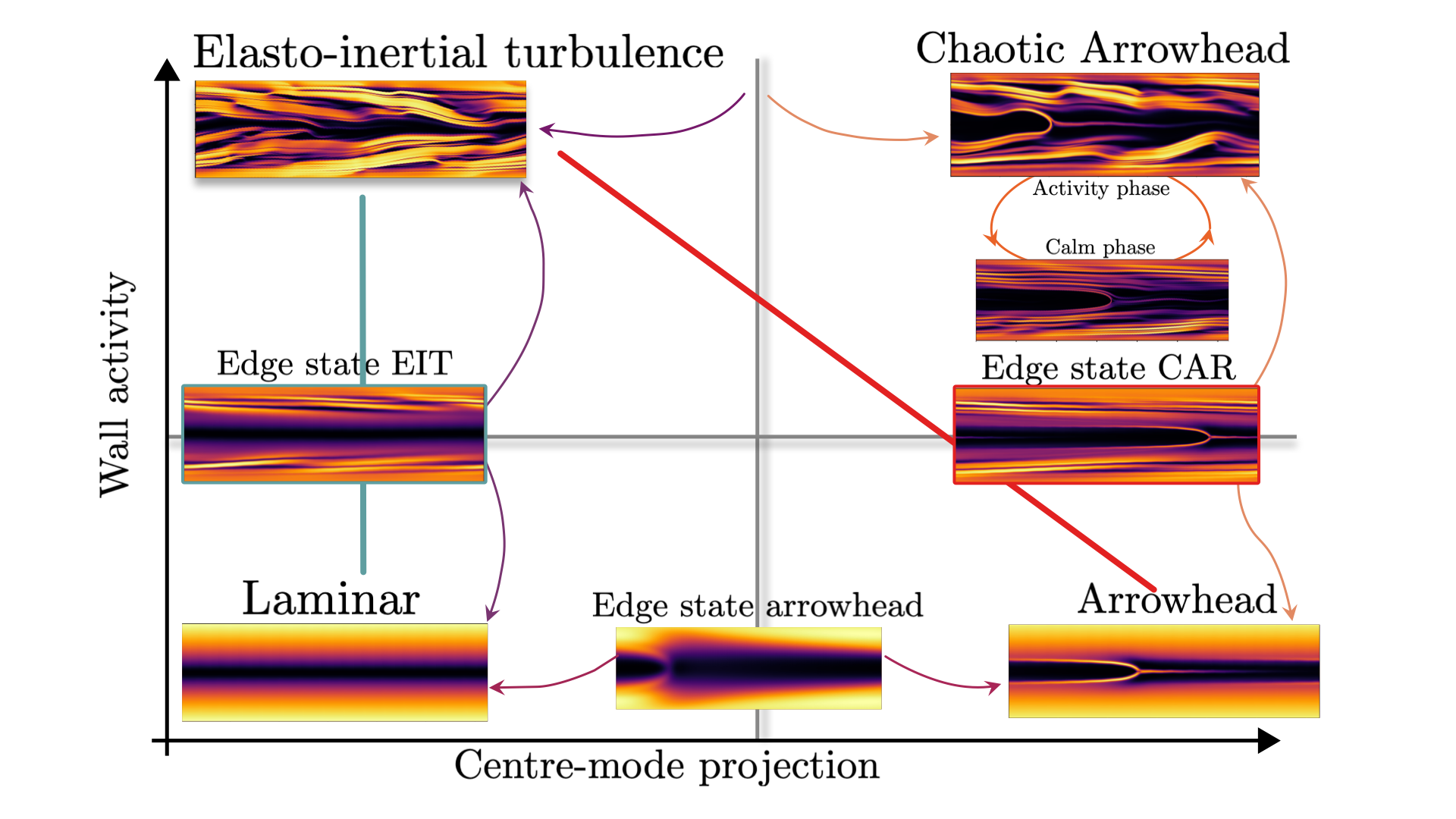}
    \caption{\MB{Sketch of the state space configuration. The four quadrants represent the basins of attraction corresponding to the states EIT, CAR, SAR, L. The solid lines emanating from the states represent trajectories approaching and departing different regions of the state space. The thick lines indicate the edge tracking carried out: between EIT and L (blue), between EIT and SAR (red) (see figure \ref{fig:edgeTrack}). The edge states resulting from the bisection algorithm are framed with the same colour. The chaotic attractors undergo calm and active phases (see figure \ref{fig:CAH_timeseries}) and approach the edge states during the calm phase.}}
    \label{fig:sketchStat}
\end{figure}

%
%
\begin{figure}
    \centering
    \begin{tabular}{cc}
        
    \includegraphics[width=0.45\textwidth]{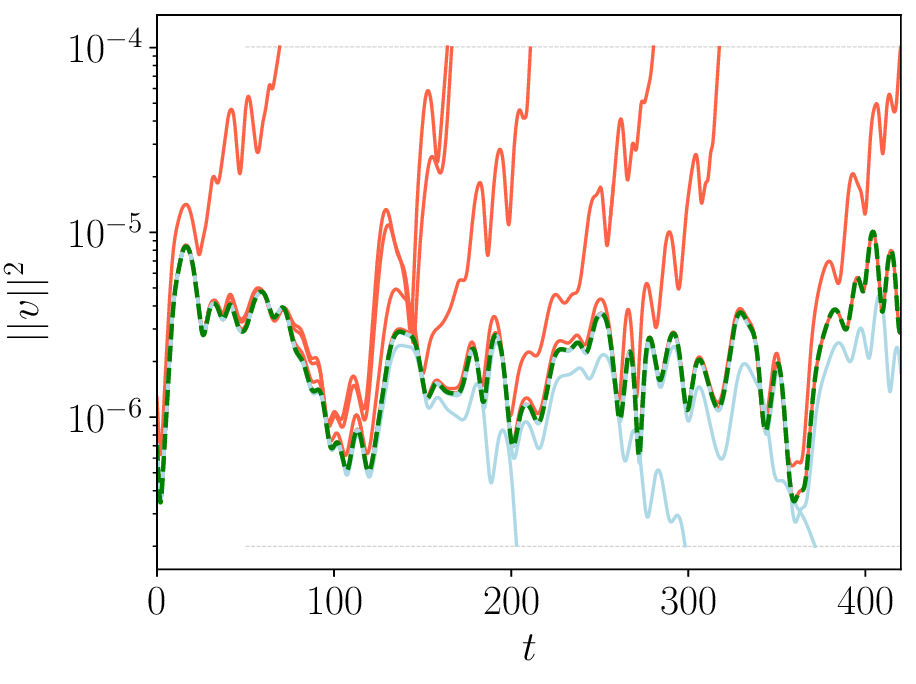}\put(-180,135){$(a)$} & \includegraphics[width=0.45\textwidth]{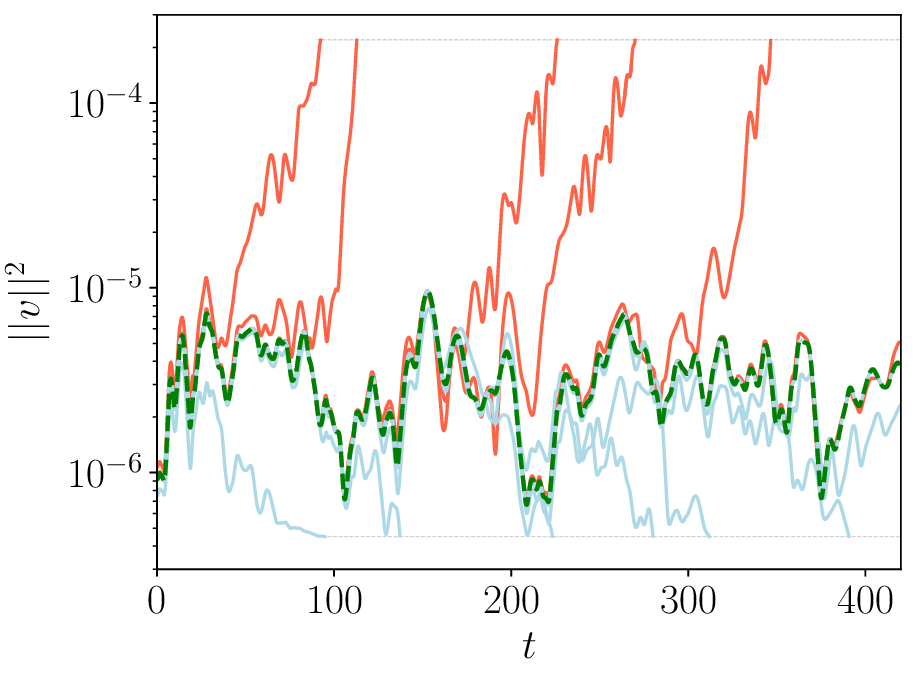 } \put(-180,135){$(b)$} \\
    (a)     & (b)\\
    \includegraphics[width=0.48\textwidth]{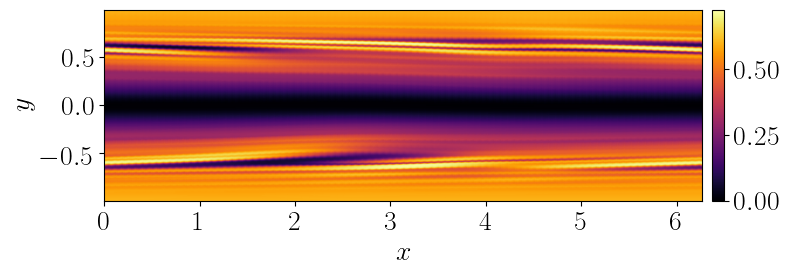} \put(-180,65){$(c)$} & \includegraphics[width=0.48\textwidth]{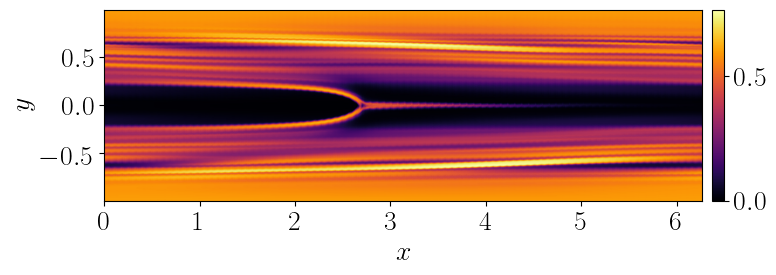} \put(-180,65){$(d)$} \\
    (c)     & (d)
    \end{tabular}
    \caption{Top edge tracking for $\textit{Re}=1000$, $\textit{Wi}=50$, $\beta=0.9$, $L_{\text{max}}=70$, $\textit{Sc}=500$: (a) between EIT and L \RK{(the green edge trajectory is bracketed by red trajectories approaching EIT  and blue trajectories relaminarising to L)}; and (b) between EIT and SAR (the green edge trajectory is bracketed by red trajectories approaching CAR instead of EIT and blue trajectories approaching SAR. 
    Bottom: \RK{(c) snapshot of tr$({\bf{C}})/{L_{\text{max}}^2}$ of the edge trajectory in (a) at $t=400$ which shows a strong polymer layer  at $y\approx \pm [0.75,0.85]$. Plot (d) repeats this for (b). Figure \ref{fig:sketchStat} explains how it is possible to reach the CAR edge state starting from a bisection between EIT and SAR}.}
    \label{fig:edgeTrack}
\end{figure}

\subsection{Edge states}

Edge states are attracting states on the edge manifold, a codimension one manifold lying in the boundary between different basins of attraction. Edge states are thus helpful to shed light on the global structure of the state space \citep{skufca2006edge,schneider2006edge,duguet2008transition}. These states can be identified by the so-called classical edge tracking algorithm based \RK{on threshold attainment of a key observable of the flow} \citep{itano2001dynamics,skufca2006edge}.

The choice of an observable to uniquely label trajectories as lying within a certain basin of attraction is not straightforward in viscoelastic flows as discussed above \RK{in \S\ref{3.2}}. The choice \RK{used here is the $L_2$-norm of the vertical velocity, $||v||^2$, which is zero for the laminar state L}. Edge tracking was then performed between (i) EIT and L \MB{(shown in figure \ref{fig:sketchStat} as a blue line)} and  (ii) EIT and SAR \MB{(shown in figure \ref{fig:sketchStat} as a red line)}.  The use of  $||v||^2$ was not able to distinguish between trajectories belonging to CAR or to EIT, as discussed in Figure~\ref{fig:statePort}. The existence of an edge manifold between CAR and EIT can still be explored by probing the state space with specific trajectories and assessing whether an arrowhead structure survives or not after sufficiently long time \RK{but this is a laborious process}.

The simplest edge state identified is the \RK{`lower branch' unstable} SAR between the \RK{`upper branch' stable} SAR and L \citep{buza2022b}. Figure~\ref{fig:edgeTrack}(a) shows the time series of the edge tracking between EIT and L and Figure~\ref{fig:edgeTrack}(c) shows a snapshot of this trajectory, a weakly chaotic state characterised by polymer layers located at $y\approx\pm [0.75,0.85]$. The edge state reveals the significance of the polymer layers located close to the walls, as they are responsible for the self-sustained chaotic dynamics within the edge. Furthermore, these layers have been observed during the calm phases of both EIT and CAR, suggesting that they related to the driving mechanism for elasto-inertial turbulence. The \RK{edge state}  between CAR and L can be compared with the \RK{edge state} found between EIT and SAR, figure~\ref{fig:edgeTrack}(b) and figure~\ref{fig:edgeTrack}(d), which also corresponds to a weakly chaotic state characterised by polymer layers located at $y\pm [0.75,0.85]$ with the presence of an arrowhead structure in the centre of the channel. 

The results of the edge tracking suggest  \RK{an organisation of  state space 
sketched in Figure~\ref{fig:sketchStat}} over a 2D plane of wall activity against centre-mode projection. Here each state is shown  within its basin of attraction and the chaotic states are shown approaching their corresponding edge states in their calm phases. \RK{Our calculations suggest that there could be an intersection between the basins of attraction of EIT, CAR, SAR and L but the saddle state residing here is not computable using bisection since it must have two unstable directions.}

 %
 %
\begin{figure}
    \centering
    \includegraphics[width=0.47\textwidth]{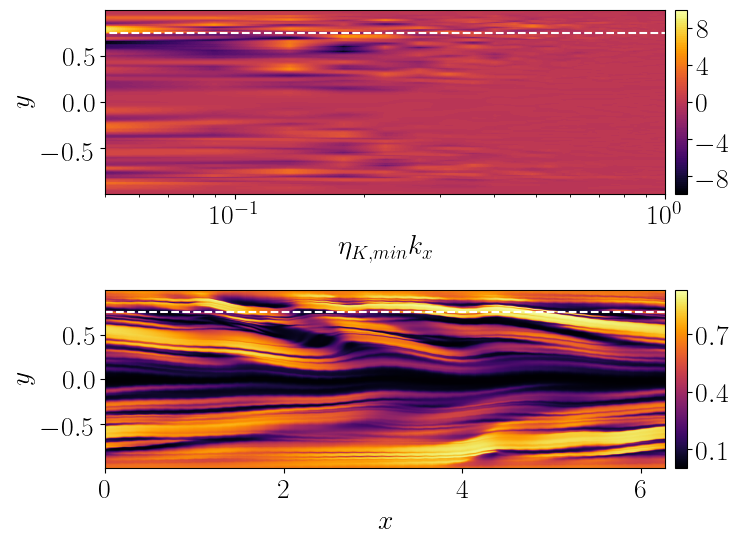}\put(-180,135){$(a)$} \put(-180,65){$(c)$}
    \includegraphics[width=0.47\textwidth]{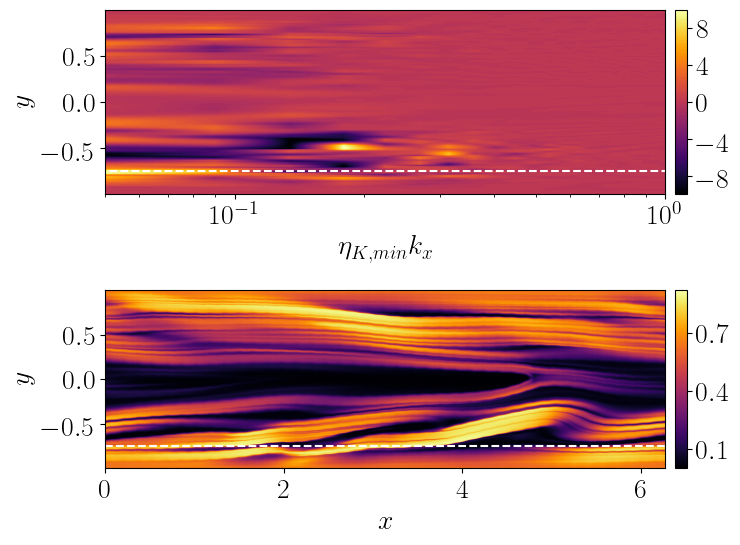}\put(-180,135){$(b)$} \put(-180,65){$(d)$} \\
    \includegraphics[width=0.47\textwidth]{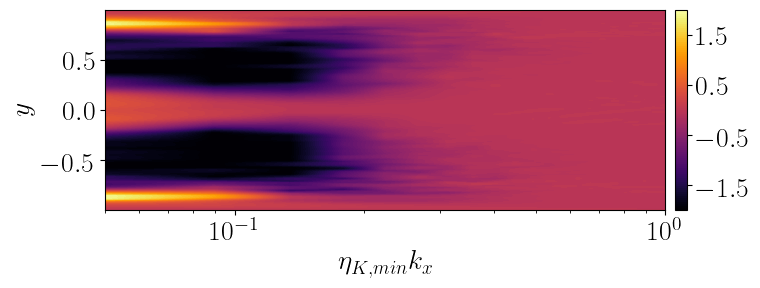}\put(-180,65){$(e)$}
    \includegraphics[width=0.47\textwidth]{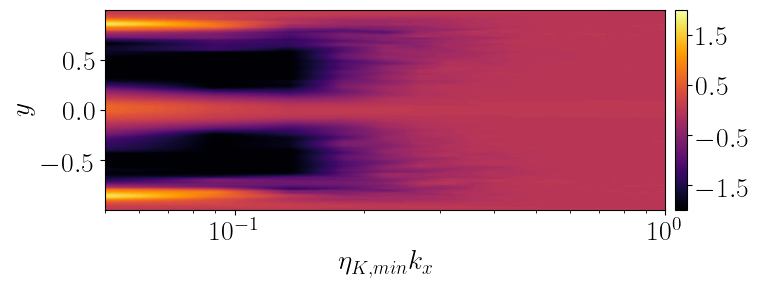}\put(-180,65){$(f)$}
    \caption{(a) Cospectra between \MB{the perturbation kinertic energy to the perturbation elastic energy} for EIT at $\textit{Re}=1000,\ \textit{Wi}=50,\ \beta=0.9,\ L_{\text{max}}=70,\ \textit{Sc}=500$ as a function of the wall-normal coordinate $y$ \MB{just before an active phase}. The streamwise wavenumber $k_x$ is normalised with the minimum mean Kolmogorov lengthscale. The white dashed line is at $y=0.75$. (c) Instantaneous snapshots of $\text{tr}(\mathbf{C})/L_{\text{max}}^2$ corresponding to the cospectra in (a). (b) same as (a) but for a snapshots of CAR at the same parameters. (d) Instantaneous snapshot of of $\text{tr}(\mathbf{C})/L_{\text{max}}^2$ corresponding to the cospectra in (b). \MB{(e) Mean cospectra for same EIT as (a). (f) Idem for CAR in (b).} \MB{The figure illustrates how the energy exchange ahead of an active phase occurs at polymer layers located at $y\approx\pm [0.75,0.85]$}.}
    \label{fig:polylayers_attractors}
\end{figure}

\RK{\subsection{Kinetic-to-elastic energy transfer}}

\RK{Our calculations of chaotic states highlight  the importance of polymer activity at the walls. Here we identify the key location where kinetic energy is transferred to elastic polymer energy. }
\RK{The energy transfer flux from} the perturbation kinetic  energy \RK{to} the perturbation elastic energy is,
\begin{equation}
    \Pi_e' := \frac{1-\beta}{\textit{Re}}T'_{ij}S'_{ij}
\end{equation}
\citep[e.g. see equations (9)-(12) in][]{dubief2020first} where primed variables indicate perturbations from the mean turbulent state, and the  dissipation rate of TKE, $\varepsilon:=\tfrac{\beta}{\textit{Re}}\partial_j u'_i\partial_j u'_i$, defines the Kolmogorov lengthscale,
\begin{equation}
    \eta_K:=\left[\frac{(\beta/Re)^{3}}{\overline{\varepsilon}}\right]^{1/4}.
\end{equation}

Figure \ref{fig:polylayers_attractors} shows the instantaneous cospectra of $\Pi'_e$ and the corresponding instantaneous field of tr$(\mathbf{C})/L_{\text{max}}^2$ \RK{for each of EIT and CAR when they are on the verge of a high-activity phase. }
These cospectra highlight that as the trajectories depart from their calm phases, the largest rate of energy exchange \RK{from} the kinetic energy \RK{to} the elastic energy, i.e. \RK{maximal $\Pi_e'$}, takes place at a location $y\approx [0.75,0.85]$ in (a) and $y\approx [-0.75,-0.85]$ in (b). This corresponds to the location of the polymer layers \MB{harvesting kinetic energy to build self-sustained chaotic dynamics.} \MB{It is also interesting to note that the main exchange from elastic energy to kinetic energy happens in the dark regions of figure~\ref{fig:polylayers_attractors}. As can be observed throughout the various figures in this work, this region supports the larger scale motions during the observed self-sustained chaotic process.}
Moreover, these polymer sheets experience the same kind of undulation for the edge states as for the complex chaotic attractors when departing from the calm phases. \RK{The time-averaged} cospectra (shown in figure \ref{fig:polylayers_attractors} (e) and (f)) confirm the importance of the energy exchange \MB{in the neighbourhood of the wall $y\approx[0.75,0.85]$}
As expected, the time-averaged cospectra for EIT and CAR are very similar as the main energy exchange driving the chaotic dynamics is located in the same region \citep{dubief2020first}.\\

%
%
\section{Discussion}


\RK{In this paper we have carried out a suite of 2-dimensional simulations of viscoelastic channel flow to explore where the various states described in \cite{dubief2020first} exist in $(Wi,Re,\beta, L_{\text{max}},Sc)$ parameter space. A fully spectral code using the FENE-P model has been used to confirm the existence of 4 distinct states: the laminar state, L, the steady arrowhead solution, SAR, a chaotic arrowhead, CAR, and elasto-inertial turbulence, EIT (the intermediate arrowhead state, IAR, of \cite{dubief2020first} has been clarified as  a CAR state where calm periods dominate over the chaotic dynamics). EIT has been found for $(Wi,Re,\beta, L_{\text{max}},Sc) \in [30,100]\times [900,1200]\times[0.9,0.97]\times [50,130] \times [500,\infty)$ with increasing $Wi$, $Re$ and $\beta$ and decreasing $L_{\text{max}}$ intensifying the chaotic behaviour. Small $Sc$ values of $\approx 50$ suppress the chaotic dynamics, while larger values of $\textit{Sc}$ allow the chaos to exist in a greater region of parameter space. }

\RK{The most significant finding, however, is that there is a substantial set of parameter values (shown in  figure \ref{fig:parReg}) where all 4 states co-exist as attractors. This contrasts with the classic `supercritical' scenario where a succession of unique attractors appear of increasing complexity as parameters are changed to make the flow more unstable (e.g. increasing $Wi$ or decreasing $L_{\text{max}}$). In particular, no evidence has been found that a bifurcation off the SAR leads ultimately to either CAR or EIT (at least in 2D) as hypothesized after the recent discovery of the centre-mode instability \citep[e.g. see][]{garg2018viscoelastic, page2020exact, Khalid2021a, shankar2022, datta2022perspectives}.
It may well be that such a subcritical bifurcation sequence exists at, for example,  higher $Wi$ or lower $L_{\text{max}}$ beyond the region of multistability. Our results do not go high enough in  $Wi$ \MB{nor low enough in $L_{\text{max}}$} to see this.
In terms of polymer concentration, SAR has been found as low as $\beta=0.5$ but remains stable even when  chaotic dynamics emerges for  $ \beta  \geq 0.9$.}

\RK{To further probe the connection between the various states, various edge states were identified between pairs of attractors,  and used to sketch the relative locations of the states in phase space. As expected, the edge state between SAR and L is the unstable `lower branch'  SAR found in \cite{buza2022a, buza2022b} while the edge states between CAR and L, and between EIT and SAR correspond to weakly chaotic states. The chaotic edge states reveal the presence of unstable polymer layers at $y \approx \pm [0.75,0.85]$, qualitatively similar to the edge states between CAR and L, and between EIT and SAR. By examining the energy transfer flux, these near-wall polymer layers were found to be where the dominant energy transfer occurs from the velocity field  to the polymers which seems fundamental for the self-sustained chaotic dynamics. In contrast the chaotic flow appeared to be \MB{insensitive} to the arrowhead structure populating the centerline region. This then further suggests that the chaotic dynamics is not related to the centre-mode instability or its arrowhead manifestation but is more  a wall-focussed phenomenon.}

\RK{The conclusion of the present study is  then that the 2D linear instability discovered by \cite{garg2018viscoelastic} in pipe flow and \cite{Khalid2021a} in a channel, and the resulting arrowhead structure \citep{page2020exact,morozov2022coherent, buza2022b}, appear dynamically disconnected from EIT at least in the 2 dimensions studied here. 
Instead, our study suggests that to trigger any chaotic motion, it is necessary to excite polymer layers located at $y\approx \pm [0.75,0.85]$ from the wall. Recent work discussing viscoelastic TS waves \citep{shekar2020self,shekar2021tollmien} suggests a plausible mechanism as  polymer stretch is found localized at the near-wall critical layer of the TS waves. Another possibility is the very recent discovery of a wall-localised linear instability \citep{beneitez2022linear}. Clearly, further efforts are needed to untangle the mechanism leading to EIT but now  this can be focussed on near-wall processes. }

\vspace{0.5cm}
Declaration of Interests. The authors report no conflict of interest.

Acknowledgements. The authors are grateful to EPSRC for supporting this work via grant EP/V027247/1. YD also thanks the support of the National Science Foundation CBET (Chemical, Bioengineering, Environmental and Transport Systems) through award 1805636.

\bibliographystyle{jfm}
\bibliography{main}

\end{document}